\documentclass[usenatbib]{flesch}

\usepackage{mathptmx}
\usepackage[T1]{fontenc}

\usepackage{graphicx}	

\title[Mars Clocks, etc.]{Mars Clocks and other novel analog clocks, using Earth Standard Seconds}

\author[E. Flesch \& R. Sanders]{\textbf{Eric Wim Flesch }$^{1}$\thanks{E-mail: eric@flesch.org}
                                and \textbf{Reggis Eric Sanders }$^{2}$  \\
\\
$^{1}$PO Box 15, Dannevirke 4942, New Zealand  \\
$^{2}$PO Box 810, Palmerston North 4440, New Zealand -- BSc Math 2024, Massey U., NZ}

\date{Accepted by Astronomical Techniques and Instruments (AstTI) 20-March-2025; received 15-Jan-2025. In original form 14-April-2024.  \\
This document is the pre-print version.}

\pubyear{2025}

\begin{document}
\label{firstpage}
\pagerange{\pageref{firstpage}--\pageref{lastpage}}
\maketitle

\begin{abstract}
We present analog clocks fitted to the Mars solar day.  These clocks use the standard Earth-based second of the International System of Units (SI) as their operational unit of time, unlike current practice for Mars timekeeping.  We discuss the importance of preserving the SI second.  On this basis, we identify the two analog clocks most suitable for public use by a future Mars population.  These are a 20-hour clock with a hand motion similar to that of the standard Earth clock, and a 24-hour clock with a novel ``Martian'' hand motion which strikes the hour when all 3 hands converge onto that hour mark on the dial.  Both clocks have Earth-day equivalents to assist learning.  We also present a 24-hour ``SpaceClock'', similar to the Martian clock but with no favored reference plane, hence equally readable from any viewing orientation.    
\end{abstract}

\begin{keywords}
Clock; Mars   
\end{keywords}


\section{Introduction}

Although there is no formal agreed system of daily timekeeping for Mars, the usual method for landed spacecraft is to stretch the hours-minutes-seconds of the 24-hour Earth clock to fit the slightly longer Mars solar day of 24 hours, 39 minutes, and 35.244 seconds (in Earth time).  Thus those Mars time-seconds have 1.0274912517 times the duration of the standard Earth seconds.  This eases the adaptation of existing software programmed in Earth time units to use Mars time units instead.  This method is used by the NASA GISS Mars24 Sunclock\footnote{\url{https://www.giss.nasa.gov/tools/mars24/}} that relies on the comprehensive analysis of Mars motion given by \cite{AM2000}.  Analog (mechanical) Mars clocks usually adhere to that schema as well, starting with the earliest version of \cite{L1954} which was powered by a ``laboratory-type heavy-duty'' motor, forward to the Mars-time wristwatches provided to the JPL team supporting the Mars Exploration Rovers ``Spirit'' and ``Opportunity''\footnote{\url{https://mars.nasa.gov/mer/spotlight/spirit/a3\_20040108.html}} in 2004.  

The notable disadvantage of that method is that the standard second, representing the physical time unit in the International System of Units (SI), is lost.  This SI second, today formally defined as the duration of 9\,192\,631\,770 periods of the unperturbed ground-state hyperfine transition of the cesium 133 atom, is a fundamental SI unit from which most other base units such as length, mass, and temperature, are derived.  It is used in all our science and technology.  Moreover, the standard second has a long association with human history: its earliest-preserved formal record dates from Ptolemy's \textit{Almagest} of the 2$^{nd}$ century CE, which described the modern division of the Earth day into 24 hours, with 60 minutes per hour and 60 seconds per minute.  Furthermore, this schema very possibly dates back 1800 years earlier to the ancient Babylon of 1600 BCE which divided its days into 24 parts and used sexagesimal (base 60) counting.  But actually throughout the ages, the time-second has always been an available measure for people in the form of the adult human heartbeat which has a time-duration comparable to the standard second.  In these ways the standard second is closely connected to our lives.

The absence of the SI second from Mars timekeeping may not be a hardship today, but in a possible future era of human-occupied cities on Mars, the people will probably not want to contend with two competing values of the ``second''.  Additionally, it would ease communications between Earth and Mars if the time-second retains its undisputed definition as that of the standard SI Earth second.  The same will likely apply to all destinations; the SI Earth second will remain an essential tool wherever we go.  

Of course, the SI Earth second does not fit exactly into Mars's solar day, so its incorporation into Mars timekeeping would require periodic temporal adjustments through use of leap-seconds or leap-hours, as well as the question of how to structure the hours-minutes-seconds.  One method found on a long-standing website\footnote{\url{http://interimm.org/mars-clock/en/timing-doc.html}} from China, consists in adding to the standard 24-hour Earth day a twenty-fifth, shorter hour containing only the remaining 39 min 35 sec of the Martian day.  However, this creates computational difficulties and cannot be used to construct simple analog clocks.  We surmise that future acceptance of a timekeeping system by the Martian public will require a day divided evenly into hours, and preferably the familiar 24 hours if possible. 

In summary, Martian timekeeping requires a simple clock, with consistent hours derived from the standard SI second.  Here we define a clock as an analog physical device, i.e., comprising a dial (clockface) with moving hands.  A useful design advantage would be if that clock design works for Earth timekeeping as well, for easy learning and conversion purposes.  In this article, we analyze which analog clock designs are thus most suitable for Mars, and we find two leading candidates which can win public acceptance.  These are a 20-hour clock and a 24-hour clock.  A variant of the 24-hour clock is shown to have appealing symmetries for space travel also.  They are presented below.

\section{Why not simply use the Earth clock with more seconds per minute?}

If it is possible to simply add more seconds per minute to the standard Earth clock to adapt it to the Mars solar day without causing undesirable complications, then this solution should be preferred.  So let's check for fit.  The Mars solar day duration is 88775.244 Earth SI seconds that we round to 88775 sec.  We now examine whether adding seconds to standard Earth minutes is sufficient to approximate the Mars solar day duration.

Using 61 sec/min, the clock day duration becomes $24\times60\times61$ = 87840 sec; this falls short of the desired 88775 sec duration by 935 sec.  Using 62 sec/min, the clock day duration becomes $24\times60\times62$ = 89280 sec, which exceeds the desired 88775 sec duration by 505 sec.  As this is the closest approximation to the Martian day length, let's suppose that it is used, in which case leap hours will be needed to keep the clock synchronized to the Mars solar day.  The length of these hours is $60\times62$ = 3270 seconds, so a leap-hour will be needed every 3270 / 505 = 7.3663366 days.  Thus, a leap hour will be needed to be subtracted from every 7-day week throughout the year (excepting about 5 weeks), so that nearly all weeks would end with a 23-hour day.  If this is publicly acceptable, then a 24hr/60min/62sec schema can work for Mars.  

If, however, it is preferred for the Mars clock day to adhere more closely to the Mars solar day, for greater consistency and that leap hours should not occur more often than daylight-savings hour shifts do on Earth, then the following sections describe the process of finding such Mars clocks, and the criteria to identify the best candidates.

\section{Considerations for an Analog Mars Clock}

The foremost consideration in designing an analog Mars clock is that it should look and function much like our own Earth clock while correctly fitting the Mars solar day.  If such a clock can be found with a perfect fit, then we will have reached our goal.  Now, the Mars solar day is 88775.244 Earth SI seconds long.  For our clock to use integer SI seconds, the excess 0.244 sec must be truncated and restored later, in a periodic leap-hour correction.  This leaves 88775 seconds to model, and if necessary we will add or subtract integer seconds from this daily amount to find the best clock; the difference will be accounted for by the necessary leap hours.

An analog dial ideally bears graduations (tick marks) which are used by all three clock hands.  The standard Earth 12-hour clock does this with 60 graduations used by both minute and second hands, with the hour hand using every fifth graduation to mark the hours.  The corresponding 24-hour clock requires 12 additional major graduations for the odd-numbered hours.  These regular graduations can be used because the SI second duration is contoured to Earth's solar day, with exactly 86400 seconds per day, by design.  Conversely, because the Earth SI second is \textsl{not} contoured to the Mars solar day, we can expect difficulties in fitting all three clock hands to a common set of graduations on the Mars clock dial.  However, the minimum requirement should be that the hour and minute hands use common graduations to preserve the clock dial coherence.  Therefore we will require that, insofar as is possible.  Also a nice-to-have would be if the Mars clock schema also works for Earth timekeeping (using a reduced number of seconds per minute), for familiarization and conversion purposes.          

On this basis, we commence our investigation at the (rounded) Mars solar day duration of 88775 SI seconds.  If we can make an analog clock with its day fitted to this optimal duration in an attractive way, our task would be done!  This number of seconds (88775) factors as $5^{2}\times53\times67$, entailing that this Mars clock would need to be modelled with 25 hours/day, 53 min/hour, and 67 sec/min.  Although this is an exact fit to the rounded Mars day duration, the resulting clock would use prime numbers of minutes and seconds, with separate graduations marked on the dial for each hand.  Such an untidy clock is not suitable for public daily use.

\begin{table} 
\centering
\caption{Prime Divisors of integers within the range 88750-88800, and clocks that can be made with such numbers of seconds per clock day.}
\begin{tabular}{rll}
\hline 
\# & divisors & comment \\
\hline
88750 & 2, 5$^{4}$, 71 & 25-hour clock, 50 min/hr ok but 71 sec/min. \\
88751 & 13, 6827 &  \\
88752 & 2$^{4}$, 3, 43$^{2}$ & 24-hour clock with 43m \& 86s or vice-versa \\
88753 & 7, 31, 409 &  \\
88754 & 2, 199, 223 &  \\ 
88755 & 3, 5, 61, 97 & 15-hour clock, 61 min/hr, 97 sec/min. \\
88756 & 2$^{2}$, 22189 &  \\
88757 & 17, 23, 227 & 23-hour clock, 17 min/hr, 227 sec/min. \\
88758 & 2, 3$^{2}$, 4931 &  \\
88759 & 11, 8069 &  \\
88760 & 2$^{3}$,5, 7, 317 & 20/28-hour clocks, but 317 sec/min. \\
88761 & 3, 29587 &  \\
88762 & 2, 44381 &  \\
88763 & 37, 2399 &  \\
88764 & 2$^{2}$, 3, 13, 569 & 26-hour clock, but 569 sec/min. \\
88765 & 5, 41, 433 &  \\
88766 & 2, 44383 &  \\
88767 & 3$^{2}$, 7, 1409 &  \\
88768 & 2$^{6}$, 19, 73 & 16/19/32-hour clocks, with 73 sec/min. \\
88769 & 29, 3061 &  \\
88770 & 2, 3, 5, 11, 269 & 15/22-hour clocks, but 269 sec/min. \\
88771 & none & prime number \\
88772 & 2$^{2}$, 22193 &  \\
88773 & 3, 127, 233 &  \\
88774 & 2, 7, 17, 373 &  \\
88775 & 5$^{2}$, 53, 67 & 25-hour clock, but 53m \& 67s, both prime. \\
88776 & 2$^{3}$, 3$^{4}$, 137 & 24-hour clock, almost perfect but see text. \\
88777 & 13, 6829 &  \\
88778 & 2, 44389 &  \\
88779 & 3, 101, 293 &  \\
88780 & 2$^{2}$, 5, 23, 193 & 20/23-hour clocks, but 193 sec/min. \\
88781 & 7, 11, 1153 &  \\
88782 & 2, 3, 14797 &  \\
88783 & 47, 1889 &  \\
88784 & 2$^{4}$, 31, 179 &  \\
88785 & 3$^{2}$, 5, 1973 & 15-hour clock, 5 min/hr, 1973 sec/min. \\
88786 & 2, 103, 431 &  \\
88787 & 19, 4673 &  \\
88788 & 2$^{2}$, 3, 7$^{2}$, 151 & 21/28-hour clocks, but 151 sec/min. \\
88789 & none & prime number \\
88790 & 2, 5, 13, 683 & 26-hour clock, 13 min/hr, 683 sec/min. \\
88791 & 3, 17, 1741 &  \\
88792 & 2$^{3}$, 11, 1009 & 22-hour clock, 4 min/hr, 1009 sec/min. \\
88793 & none & prime number \\
88794 & 2, 3$^{2}$, 4933 &  \\
88795 & 5, 7, 43, 59 & good luck with this one (35h/43m/59s). \\
88796 & 2$^{2}$, 79, 281 &  \\
88797 & 3, 29599 &  \\
88798 & 2, 29, 1531 &  \\
88799 & none & prime number \\
88800 & 2$^{5}$, 3, 5$^{2}$, 37 & 20/24/25-hour clocks with 50/60 min/hr \& \\
      &                         & 74 sec/min, or 100 min/hr with 37 sec/min. \\
\hline
\multicolumn{3}{l}{Note: Divisors reveal the available timekeeping patterns applicable to Mars} \\
\multicolumn{3}{l}{clock designs. Example calculation: The top row gives the prime divisors} \\
\multicolumn{3}{l}{of 2,5,5,5,5, and 71.  These are combined to obtain $5\times5$ = 25 hours/day,} \\
\multicolumn{3}{l}{$2\times5\times5$ =50 min/hr, and the remainder (71) provides the sec/min.} \\
\multicolumn{3}{l}{In this case, the 71 sec/min is considered undesirable for clock design,} \\
\multicolumn{3}{l}{because 71 is a large prime number, thus indivisible.} \\
\end{tabular}
\end{table}

Accordingly, we add or subtract integer seconds from the 88775 seconds duration, in search of a ``clock day'' duration yielding a suitable analog clock within a $\pm$25 seconds offset, i.e., within the range 88750-88800 sec.  Table 1 lists all these, showing the number of integer SI seconds for each proposed clock day, its prime-number divisors, and a comment to highlight any interesting feature (the table footnote explains how to combine the divisors into hours, minutes, and seconds).  Most of these numbers of seconds have large prime numbers as factors, thus cannot be used.  Interesting rows are:  

\begin{itemize}
\item A clock day of 88750 sec factors as $2\times5^{4}\times71$ which (as the Table 1 footnote explains) yields 25 hours/day, with 50 min/hr and 71 sec/min (such clock schema hereinafter written as, in this case, 25h/50m/71s).  For this clock, the hour and minute hands both use the same 50 dial graduations which is a positive aspect; however, the 71 sec/min is a somewhat large prime number, indivisible and thus unfriendly for public use.  Also the 25 hours/day is not evenly divisible except by 5, and this clock cannot be used for Earth.  Still, such a clock would be useable.  
\item A clock day of 88752 sec yields a 24-hour clock as 24h/43m/86s or 24h/86m/43s.  Here the minute and second hands both use the same 86 graduations on the dial, but the hour hand requires a completely different set of 24 graduations.  Moreover, the 43 min/hr (or sec/min) is a prime number, and this clock also has no Earth equivalent.  This 24-hour clock could be constructed, but would be impractical for regular daily use.
\item A clock day of 88776 sec also yields a 24-hour clock, but the prime divisor of 137 would entail use of 137 sec/min and only 27 min/hr.  This could not be popular.  However, this 88776 sec duration is offset only 0.756 sec from the Mars solar day of 88775.244 sec.  This could be highly suitable if 137 could be divided into equal halves: then, the resulting timekeeping pattern would be 24h/54m/68.5s with no need for a leap-hour adjustment for decades!  Unfortunately, the corresponding analog clock dial would have no coherence, with separate graduations for each hand.  Moreover, there is no easy solution to manage the half-second that this pattern would generate.  With regret we must reject this schema.   
\item A clock day of 88800 sec factors as $2^{5}\times3\times5^{2}\times37$.  Now this offers us many combinations: 20h/60m/74s, 24h/50m/74s, 24h/100m/37s, 25h/48m/74s, 25h/96m/37s, 15h/80m/74s, and 16h/75m/74s.  The 20-hour clock looks a strong candidate for public acceptance, with both hour and minute hands using 60 graduations on the dial, with the familiar 60 minutes per hour as on Earth (albeit with only 20 hours in the day), and the 74 sec/min is divisible by 2.  Also it has an Earth-equivalent version with 72 sec/min, as $20\times60\times72$ = 86400 sec/day which is the Earth solar day duration.  The 24-hour clocks also have attractive minute sizes but we would not expect to be able to use the same dial graduations for both the 24 hours and the 50 (or 100) minutes.  But in fact this is found to be possible by using a novel ``Martian'' hand motion described in Section 5.1.  These 24-hour clocks also have equivalent Earth versions, with $24\times50\times72$ or $24\times100\times36$ schemas both yielding 86400 sec/day.
\end{itemize}
 
The end result of this survey of possible timekeeping schemas is that using 88800 SI seconds for an operational Mars clock day enables the design of practical and interesting candidate analog clocks which could be accepted for daily use by a future Martian public.  Discounting the 25hr/day clocks which are unlikely to be popular due to the lack of even divisibility of hours within the day, the best candidate clocks are seen to be the 20-hour clock with standard hand motion, and the 24-hour clocks with an original hand motion described in Section 5.1. The leap-seconds needed to keep these 24-hour clocks aligned with Mars's rotation come to 24.756 seconds per Mars solar day, which accumulates to the need for 4 or 5 leap-hour shifts per Martian tropical year, a frequency comparable ($\approx{18.9}\%$ more often, measured in SI seconds) with that of the twice-yearly daylight savings hour shifts on Earth. 

The 88800 SI seconds per day used for these Mars clocks, amount to exactly 37/36 of the 86400 seconds of the Earth day.  This exact ratio means that any Mars clock using a multiple of 37 sec/min has an equivalent Earth clock with 36 sec/min (or that same multiple thereof), all else being unchanged.  Therefore, equivalent Earth clocks can be derived from all of these 88800 sec/day Mars clock schemas because they all use 37 or 74 (=2$\times$37) sec/min.

\section{The 20-hour Mars clock}

Using 88800 SI seconds as the operational Mars day, the 20-hour clock described in Section 3 is the most straightforward schema.  Its 20h/60m/74s timekeeping pattern, allowing for many useful subdivisions of the day and hours, is friendly for public use.  Other attractions are its standard hand motion and the familiar 60 minutes/hour.  People can likely become accustomed to the 20-hour day and the 4440-sec long hours.  The leap-hour shift would be a larger adjustment to the day, being 4440 seconds, but would accordingly be required less often, at approximately the same frequency (0.9\% less often, measured in SI seconds) as the twice-yearly daylight savings hour shifts on Earth.  Furthermore, this 20h/60m/74s schema allows for a 10-hour AM/PM analog clock, similar to standard Earth clocks.  Figure 1 shows the two clockfaces with an example time.  These clocks also works for Earth with 72 seconds per minute, but realistically would have no actual use except perhaps as an aid to converting Mars time to Earth time.  Nonetheless the 20-hour clock can be seen showing your local Earth time (i.e., the reader's computer clock time), at \url{https://quasars.org/clock/marsclock-for-earth.htm?mode=20} (or specify mode=10 for the 10-hour clock) as an analog clock with the equivalent digital display underneath, and the standard Earth time on the browser tab above.

\begin{figure} 
\includegraphics[scale=0.22, angle=0]{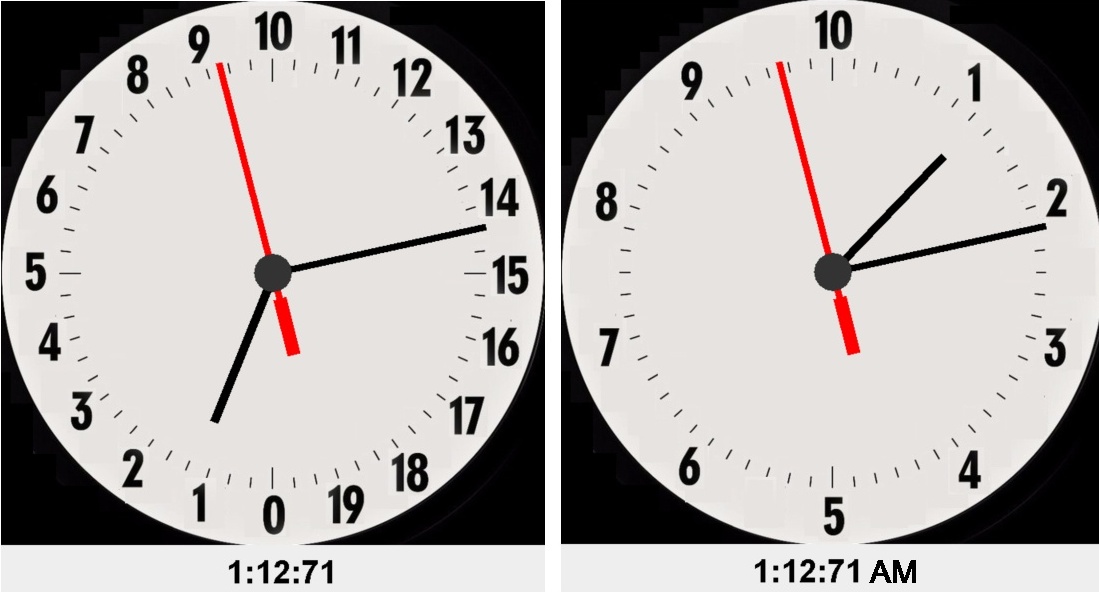} 
\caption{The proposed 20-hour Mars analog clock, with its 10-hour AM-PM variant.} 
\end{figure}

\section{The 24-hour Mars Clocks}

Because the Mars solar day is, at 24.65979 Earth hours, similar in duration to the Earth day, there is a general expectation that the incipient Mars clock will also feature 24 hours in the Mars day, for familiarity and comfort.  This is easily done with digital clocks: dividing the exact Mars day duration (88775.244 SI seconds) into 24 hours yields 3698.9685 seconds per hour, which is brought up to an even 3700 sec/hr by our adoption of the operative clock day of 88800 SI seconds duration (Section 3).  Therefore, such a digital clock can trivially display 24h/50m/74s or 24h/100m/37s timekeeping patterns, also in a 12-hour AM/PM format if desired.  Note that the 24h/100m/37s schema amounts to displaying only decimal hours (to the hundredth) and seconds; such a digital clock could display as \textsl{hh.hh:ss}, omitting minutes altogether.  If minutes are preferred as a distinct time entity, then the 24h/50m/74s schema can be used.  

This digital clock can also display Earth time using 36 sec/min instead of 37 sec/min, e.g., using the 24h/100m/36s Earth schema, 12.24:35 on this clock would be equivalent to 12:14:59 on the standard Earth clock, both showing a time 1 second before a quarter-hour past 12:00.

\subsection{The 24-hour Analog ``Martian'' Clock}
         
The corresponding analog 24-hour clock (24h/100m/37s variety) requires marking 100 graduations for minutes onto a 24-h dial, which seems incompatible.  However, this is accomplished by inserting 3 minor graduations between the major (hour) graduations for a total of 96 graduations on the dial; moreover, because the hour hand progresses over 4 graduations per each hour -- thus the minute hand, starting each new hour upon passing over the hour hand, then completes a full rotation of 96 graduations (minutes) and then progresses 4 more graduations to catch up to the hour hand, thus completing the required 100 minutes.  Figure 2 gives three examples of the exact hour striking when the minute hand overlaps (i.e., passes over) the hour hand.  Similarly, the second hand passes over the minute hand at the stroke of each exact minute, including that minute which coincides with the stroke of the hour.  Therefore at the stroke of each hour, all three hands meet at that hour mark on the dial.  We label this original clock movement as the ``Martian'' movement, achieved with the following periodicity of the clock hands for each clock day: 

\begin{itemize}
\item The hour hand performs one full rotation around the dial.  
\item The minute hand performs 25 full rotations around the dial.
\item The second hand performs 2425 full rotations around the dial for 100 min/hr, or 1225 full rotations for 50 min/hr.
\end{itemize}

\begin{figure} 
\includegraphics[scale=0.18, angle=0]{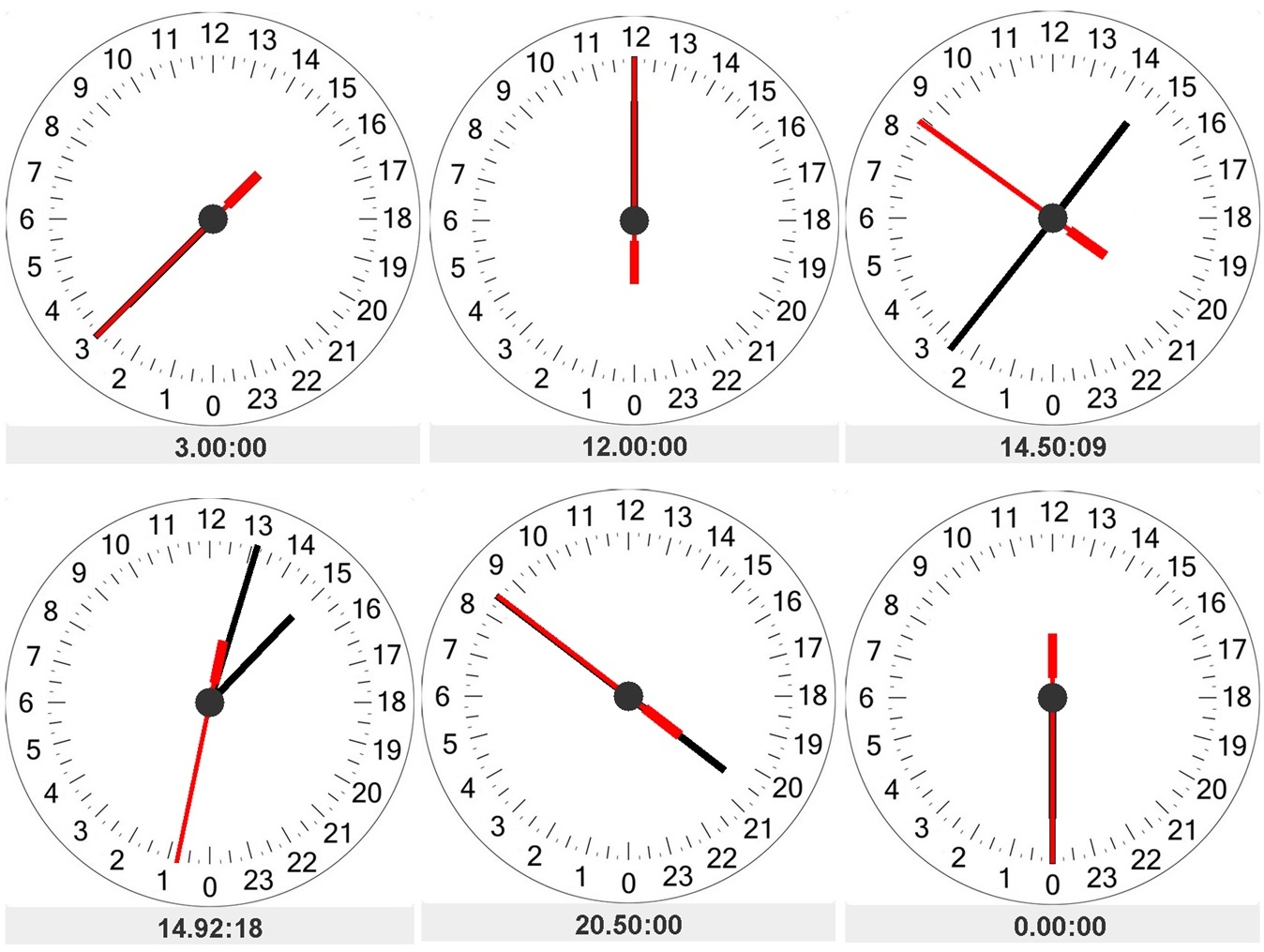} 
\caption{The Martian analog clock of 24h/100m/37s (this dial called the ``exact'' Martian clock) for Mars time; a selection of times shown, with digital time beneath each panel; the digital times have the pattern ``hh.hh:ss'', thus display as hours, hundredths-of-hours, and seconds (the ``minute'' serves as an alias for 1/100$^{th}$ hour).  Top left: 3 o'clock: all 3 hands point to the ``3'' on the dial.  Top center: Noon, i.e., 12:00: all 3 hands point to the ``12'' on the dial.  These are instances of the rule that at every exact hour, all 3 hands meet at that hour graduation on the dial.  The minute hand uses the hour hand as its reference, thus its digital value is zero when crossing over the hour hand, and 99 just prior.  The second hand uses the minute hand as its reference, thus its digital value is zero when over the minute hand, and 36 just prior.  Bottom center: 20.5 hours; at the exact half-hour, the minute hand points in the opposite direction from the hour hand, and the second hand crosses over the minute hand on the stroke of the minute.  Bottom right: Midnight.} 
\end{figure}

\begin{figure*} 
\includegraphics[scale=0.3, angle=0]{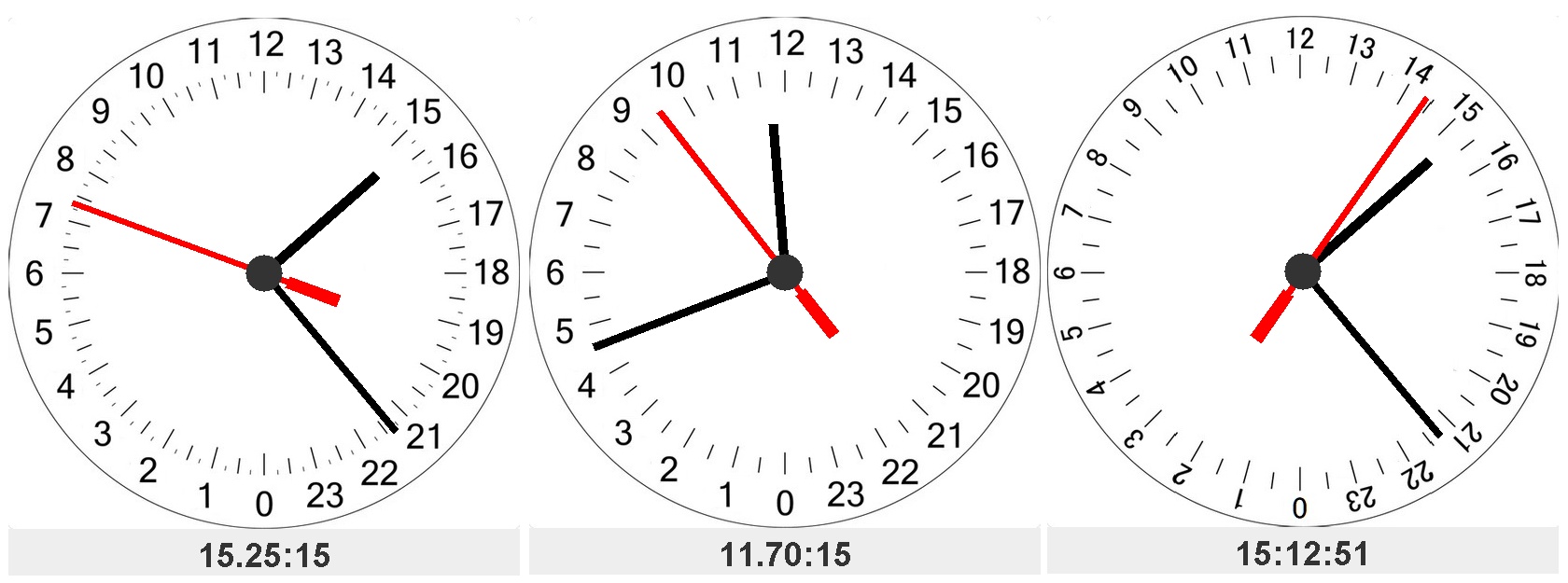} 
\caption{The three desktop clock faces which use the ``Martian'' hand motion.  Left: the ``exact'' Martian clock, the second hand sweeps once per each hundredth-of-an-hour (i.e., a short minute: 37 seconds on Mars or 36 seconds on Earth); Middle: the ``relaxed'' Martian clock, the second hand sweeps once per each fiftieth-of-an-hour (i.e., a long minute: 74 seconds on Mars or 72 seconds on Earth), used either as one long or two short minutes; Right: the ``SpaceClock'', for free-floating crew in spacecraft, the second hand sweeps once per each long minute.  The left and right clocks show the same Earth time, 915 seconds past 15 o'clock, expressed by the left clock in short minutes, and by the right clock in long minutes.} 
\end{figure*}

This clock can also display Earth time by using 36 sec/min instead of the 37 for Mars; an online version displaying your local Earth time (i.e., the reader's computer clock time) can be viewed at 
\url{https://quasars.org/clock/marsclock-for-earth.htm?mode=e} with the digital time underneath, and the standard Earth time on the browser tab above.  The rapid rotation of the second hand around this dial may not have eye appeal to some people. The alternative 50 min/hr schema provides a leisurely 74 seconds (for Mars; 72 seconds for Earth) for a single sweep of the second hand around the dial (hereinafter for the Martian clock, a ``sweep'' of the second hand is defined as its rotation around the dial from one crossing over the minute hand to the next such crossing; similarly, a sweep of the minute hand consists of its rotation around the dial from one crossing over the hour hand to the next such crossing). Note that this 50 sweeps/hr schema requires only 48 graduations on the dial, so it is a pleasantly sparse clock face which we dub the “relaxed” Martian clock\footnote{The 50 min/hr relaxed Martian clock displays your local Earth time (i.e., the reader's computer clock time) at \url{https://quasars.org/clock/marsclock-for-earth.htm?mode=r&digital=s} with the digital time underneath, and the standard Earth time on the browser tab above.}.  This dial can also be used for the 24h/100m/37s schema (or 36s for Earth) if a single sweep of the second hand accounts for 2 minutes, although it is less intuitive at first, because the minute is struck also when the second hand points in the opposite direction from the minute hand\footnote{The 100 min/hr relaxed Martian clock displays your local Earth time (i.e., the reader's computer clock time) at \url{https://quasars.org/clock/marsclock-for-earth.htm?mode=r} with the digital time underneath, and the standard Earth time on the browser tab above.}.  The second hand does not relate to any dial graduations, but ticks the seconds independently. Figure 2 shows six times of a Martian day depicted on the analog Martian clock, starting with 03:00 when all 3 hands point to the ``3''; the dial places the midnight hour at bottom and 12-noon at top, so that the hour hand follows the Sun’s apparent course in the sky (i.e., its azimuthal angle).


Note that the 12$^{th}$ hour on the Martian dial is simply one of the 24 hours, with no positional significance, contrary to standard Earth clocks where it serves as the anchor (zero value) for the minute and second hands. On a standard Earth clock, the three hands move independently, whereas on the Martian clock the three hands move interconnectedly, with their relative positions indicating the time. To tell the Martian time at a glance, note the hour by locating the hour hand, then note the orientation of the minute hand relative to the hour hand, which corresponds to the orientation of the minute hand to the ``12'' graduation on the standard 12-hour Earth clock. Thus, at the half-hour, the minute hand points in the opposite direction from the hour hand, as it does from the ``12'' graduation on a standard Earth clock. The second hand is similarly oriented relatively to the minute hand, pointing in the opposite direction at the half-minute. The three dials designed with the Martian hand motion and described in this article are shown in Figure 3.

No 12-hour AM–PM analog Martian clock is available, because common clock graduations cannot be used for both the hour and minute hands. This is not a problem for a digital clock, so that fabricating a 12-hour AM–PM digital display for the Martian clock is trivial.
   
An analog Martian Clock application that we have designed for computer desktops is available for download\footnote{Computer desktop application available at \url{https://quasars.org/Martian-Clock.zip}}; it is initially set to display Earth local time, with 36 seconds per sweep of the second hand, and a digital display underneath it. Other clock hand motions and dials are available, including whether the clock is to be for Mars or Earth, or large or small displays; 42 preset presentations are available, and many more options for those wishing to manually tweak the ``settings'' file. The downloadable zip file includes a ReadMe text file detailing the optional settings. A Java 7 (or higher) runtime environment is required to run the clock.

This novel analog hand motion could be seen as an alternative to the standard hand motion of today. It was originally developed as a 24h/50m/72s Earth clock in 2015 and put on-line as a ``SpaceClock'' in 2019. Its main attraction was the novel hand motion, but also featured optional ``gravity'' and ``orbital'' dials. The ``gravity'' version (with upright hour numbers) evolved into the Relaxed Martian 24h/50m/72s clock presented here, after we determined its serendipitous suitability to the Martian solar day. The ``orbital'' version is the SpaceClock displayed in Figure 3, which has no preferred reference frame (because the ``12'' graduation is not an anchor for any clock hand, and the hour numbers rotate with the dial edge) and is thus equally readable from any viewing orientation; it could be suitable as a functional ornament for free- floating spacecraft crew\footnote{The SpaceClock displays your local time (i.e., the reader's computer clock time) at \url{https://quasars.org/clock/marsclock-for-earth.htm?mode=s} with the digital time underneath, and the standard Earth time on the browser tab above.}.

\begin{figure*} 
\includegraphics[scale=0.35, angle=0]{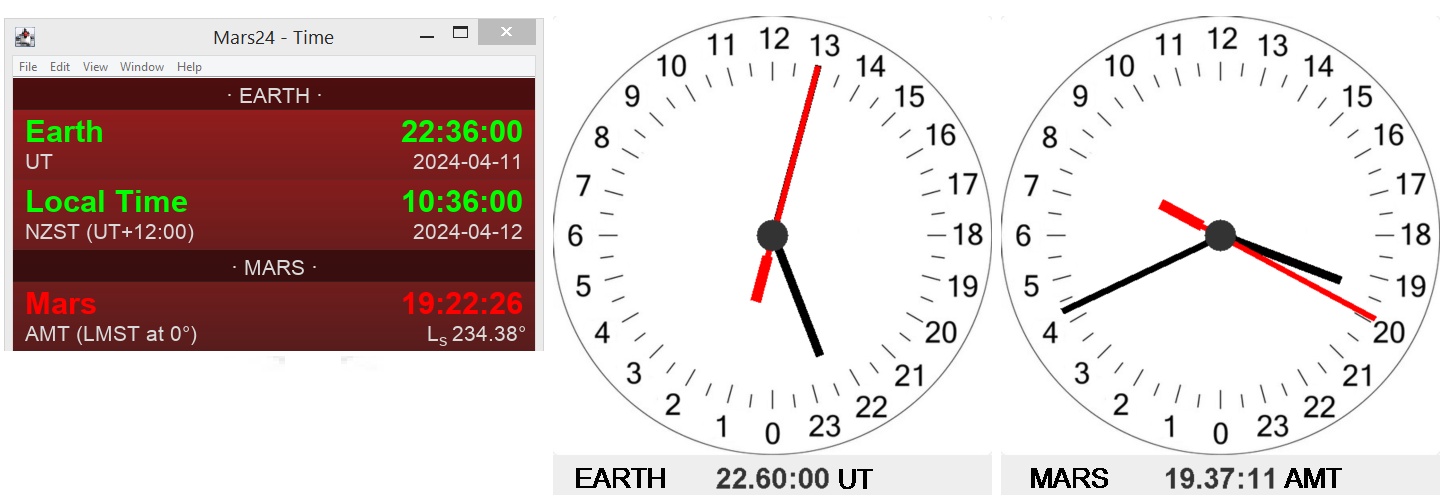} 
\caption{NASA Mars24 time compared with our proposed Martian clock times.  Earth UT of 22:36 is expressed by the relaxed Martian clock (Earth version) as 22.6 hours, with the second hand overlapping the minute hand at the stroke of that minute.  Mars24 AMT of 19:22:26 is expressed by the relaxed Martian clock (Mars version, counting two short minutes per sweep of the second hand) as 19.37 hours and 11 seconds, i.e., 16 seconds ahead of the displayed Mars24 time as discussed in the text.} 
\end{figure*}

\section{Comparison of the Martian Clock to the NASA Mars24 Clock}

The NASA Mars24 clock\footnote{available at \url{https://www.giss.nasa.gov/tools/mars24/}} presents the Mars timekeeping system used by most Mars landers; its daily clock consists of a variant of Earth time with hours. minutes, and seconds all ``stretched'' by a multiplier of 1.02749125 to fit the Martian solar day which is longer than the Earth solar day by that ratio.  In so doing it loses the standard SI second, of course, which is what the Martian clock, presented in this paper, is designed to remedy.  So it would be beneficial to reconcile these two timekeeping systems to verify that the Martian clock is keeping Mars time accurately, i.e., in accordance to current benchmarks.

Figure 4 shows a photograph in time of the Mars24 display alongside two working demonstration Martian clocks for Earth UT and Mars AMT (available from the downloadable zip file referenced in footnote 6). The caption of Figure 4 details the matches, but in brief the Earth UT (Universal Time, aka Greenwich Mean time) times match exactly (as they trivially should), but the Mars AMT (Airy Mean Time, the UT of Mars) times are offset by 16 seconds. This is seen by converting both times to SI seconds elapsed since midnight; first, the Mars24 AMT of 19:22:26 translates to 69746 Mars24-``stretched'' seconds, which multiplied by the stretch factor of 1.0274912517 come to 71663.4 SI seconds since midnight. Next, the Martian time (also AMT) of 19.37 hours and 11 seconds, with each hour comprising 3700 SI seconds, come to 71680 SI seconds since midnight. Thus there is an offset of 16.6 seconds between the two clocks. This appears to be a consistent zero-point offset between the Martian and Mars24 midnights, as follows:

First, by design, our Martian clock uses an operational Mars day of 88800 SI seconds, hence drifts from the true Mars solar day of 88775.244 SI seconds; it requires periodic use of leap hours, approximately four times per tropical Mars year. However, for this demonstration model, we reinitialize the Martian clock after 88775 (or 88776) SI seconds to keep its days aligned with the true Mars solar day. This is done using the method of \cite{AM2000} which gives 6 January 2000 (00:00:00 UTC) as the moment at which the Mars AMT and Earth UT midnights coincide (hereinafter “zero-time”). Therefore for any queried Earth date-time, we calculate the corresponding offset from the Earth zero-time in SI seconds and then add that to the Mars zero-time, incrementing the Mars solar day count by 1 per every 88775.244 SI seconds. In this way we arrive at the Mars date and time corresponding to the initial query.

The Mars24 calculational method is described online at \url{https://www.giss.nasa.gov/tools/mars24/help/algorithm.html}, where in their Section II ``Worked Examples'', they describe their Mars zero-time to be 21 Mars-seconds (i.e., approximately 21.577 SI seconds) \textsl{after} that of \cite{AM2000}.  Furthermore, they include 6 leap Earth seconds in their calculation of the elapsed duration since the Earth zero-time; if reported onto the Earth zero-time (effectively advancing it by 6 sec), this is equivalent to reducing their zero-time offset between Earth-time and Mars-time to approximately 15 seconds, comparable to the offset of 16.6 seconds that we are investigating.  This confirms that our Martian clock, by running approximately 16 seconds ahead of the Mars24 clock, is correctly keeping Mars time as per the schedule set for it.          

Also, the Mars24 clock shows the Earth \& Mars clocks ticking off the seconds at arbitrary offsets to each other, which of course happens because those seconds are of different durations.  By contrast, because the Martian clock uses standard SI seconds, it is of course set to tick simultaneously with Earth’s ticks (by whichever criteria establish simultaneity).

\section{Conclusion} 

Clocks for Mars which preserve the standard SI second are evaluated.  We select 88800 SI seconds as the best operative Mars clock day duration with which to make simple analog Mars clocks.  Using that, we demonstrate that the best (i.e., most likely popular) available clocks are a 20-hour clock with the standard hand motion of Earth clocks, and a 24-hour clock with a novel ``Martian'' hand motion which strikes the hour when all 3 clock hands point to that hour.  The 24-hour clock comes in 3 potentially useful variants.  A 10-hour analog AM–PM clock can be made for the 20-hour clock, but a 12-hour analog AM–PM clock cannot be made for the 24-hour clock.  Digital AM–PM clocks for both models are trivially makeable.  A leap hour will be needed about four times per Mars year, because the Mars solar day is $\sim$24 seconds shorter than this operational Mars clock day.  Also an analog ``SpaceClock'' is presented for space travel.  All these clocks can also keep exact Earth time.

\section*{Data Availability}
 
A computer desktop application which enables display of all these clocks and their variants is available for download at \url{https://quasars.org/Martian-Clock.zip}.  It requires a Java 7 (or better) runtime environment.

\section*{Acknowledgements}
This work was not funded.

\section*{Abbreviations List}
\begin{itemize}
\item  AMT: Airy Mean Time (the UT of Mars). 
\item  GISS: Goddard Institute for Space Studies (within NASA).
\item  JPL: Jet Propulsion Laboratory (a division of NASA).
\item  NASA: National Aeronautics and Space Administration (USA).  
\item  UT: Universal Time (aka Greenwich Mean Time).
\item  UTC: Coordinated Universal Time: like UT, but measured with atomic clocks, with leap seconds used to keep aligned to UT.
\end{itemize}

\section*{Author Contributions}
Eric Flesch conceived the Martian Clock hand movement, and wrote this manuscript.  Reggis Sanders wrote the Java and JavaScript programs which have animated these clocks and made them available for public download.

\section*{Declaration of Interests}
The authors declare no conflicts of interest.


\bsp	
\label{lastpage}
\end{document}